# In-operando synchronous time-multiplexed O K-edge x-ray absorption spectromicroscopy of functioning tantalum oxide memristors

Suhas Kumar,[1,2] Catherine E. Graves,[1] John Paul Strachan,[1,a] A. L. David Kilcoyne,[3] Tolek Tyliszczak,[3] Yoshio Nishi[2] and R. Stanley Williams[1]

[1] *Hewlett Packard Laboratories, 1501 Page Mill Road, Palo Alto, California, 94304, USA,*
[2] *Department of Electrical Engineering, Stanford University, Stanford, California, 94305, USA,*
[3] *Advanced Light source, Lawrence Berkeley National Laboratory, Berkeley, California, 94720, USA*

[a] Electronic mail: John-Paul.Strachan@hp.com

Memristors are receiving keen interest because of their potential varied applications and promising large-scale information storage capabilities. Tantalum oxide is a memristive material that has shown promise for high-performance nonvolatile computer memory. The microphysics has been elusive because of the small scale and subtle physical changes that accompany conductance switching. In this study, we probed the atomic composition, local chemistry and electronic structure of functioning tantalum oxide memristors through spatially mapped O K-edge x-ray absorption. We developed a time-multiplexed spectromicroscopy technique to enhance the weak and possibly localized oxide modifications with spatial and spectral resolutions of <30 nm and 70 meV, respectively. During the initial stages of conductance switching of a micrometer sized crosspoint device, the spectral changes were uniform within the spatial resolution of our technique. When the device was further driven with millions of high voltage-pulse cycles, we observed lateral motion and separation of ~100 nm-scale agglomerates of both oxygen interstitials and vacancies. We also demonstrate a unique capability of this technique by identifying the relaxation behavior in the material during electrical stimuli by identifying electric field driven changes with varying pulse widths. In addition, we show that changes to the material can be localized to a spatial region by modifying its topography or uniformity, as against spatially uniform changes observed here during memristive switching. The goal of this report is to introduce the capability of time-multiplexed x-ray spectromicroscopy in studying weak-signal transitions in inhomogeneous media through the example of the operation and temporal evolution of a memristor.

**I. INTRODUCTION**

Several materials, especially some metal oxides, exhibit memristive behavior involving non-volatile conductance switching when driven by an electric potential or current. This effect can be used to store information and is a frontrunner for next generation storage-class memory owing to its low power, scalability, high speed, long retention and high endurance relative to competing technologies.[1-4] Memristive materials have also received attention because of their applicability in neuromorphic computing, micromechanics, optical switching and sensing.[5-8] Tantalum oxide in particular shows promise for especially high endurance and low power storage.[9-12] Despite the level of activity, the microphysics of their operation is still only partially understood.[13] The accepted model for conductance switching is the existence of nanoscale conduction channels that form and break because of oxygen ion or vacancy migration upon application of electrical stimuli.[9,12,14-17] Some of the uncertain details of the switching mechanism in tantalum oxide involve the chemical nature of the conduction channel, the effect of the film stoichiometry on the device performance, the roles of Joule heating and electric field in inducing the switching and the extent of localization and cyclability of switching in a channel.

We employed in-operando x-ray absorption spectromicroscopy to study the local chemical, structural and electronic changes in functioning crosspoint tantalum oxide memristors. To enhance the weak signals and eliminate long-time drift when collecting data, we utilized continuous time-multiplexing. We observed uniform changes in the x-ray absorption in the film throughout the crosspoint area (2 μm x 2 μm) as the device was switched conventionally between conductance states within the resolution of the technique (~30 nm). We then studied the effects of electric field and Joule heating within the device. After further cycling a device with millions of high voltage pulses, we observed agglomerates of oxygen rich and oxygen deficient spatial regions with size on the order of 100 nm scattered over the crosspoint. We use the spatial and spectral data to better understand the relation of the material behavior to the lifecycle of a memristor. We also discuss some engineering implications of our findings.

**II. EXPERIMENTAL PROCEDURE**

X-ray absorption spectromicroscopy gives information on the unoccupied partial density of states accessed by exciting electrons from occupied levels. Here we used the oxygen 1s electrons (O K-edge) to probe the unoccupied conduction band density of states of the tantalum oxide film. High intensity x-rays from the Berkeley ALS synchrotron were filtered in energy to a spectral resolution of about 70 meV and focused to a spatial resolution down to 30 nm, while imaging was enabled by precisely controlled stage movement.[18,19] The crosspoint devices measuring 2 μm x 2 μm in area were fabricated on 200 nm low stress silicon nitride membranes that offer a low cross section for x-ray absorption. The device consisted of two platinum electrodes (20 nm thick) that formed the crosspoint across a blanket of 6 nm of tantalum oxide ($TaO_x$) and 7 nm of patterned tantalum metal electrode above it, giving the final stack (bottom to top): $Pt/TaO_x/Ta/Pt$.[11] The device was wire-bonded for in-operando x-ray measurement in the spectromicroscope and could be switched between conductance states using voltage pulses or DC voltage/current sweeps. All voltages were applied to the top electrode relative to the grounded bottom electrode.

The spectral changes upon switching are expected to be extremely subtle and possibly localized, leading to a low signal to background ratio. Because of drift in the spatial position of the sample, changes in the background absorption, variations in the x-ray intensity and possible dynamics in the system,[20] it is challenging to isolate the signal corresponding to the conductance state of the device[9,16,21] (also see supplemental material[22], Figure S1). To overcome this problem, we switched the device under study between two states at speeds on the order of a few kHz while x-ray data was acquired spatially, spectrally or both (Figure 1a). The two induced states could be high-low conductance, hot-cold (Joule heating cycling) or positive-negative electric field. The x-ray detector signal was synchronously gated into two different counters corresponding to the two repeatedly induced states of the device (continuous time-multiplexing). Thus, data at every position in a spatial map or energy in a spectral scan consists of several coincident measurements of the two states, thereby amplifying the differences at every data point. This technique reduced the time delay between corresponding pixels of an ON-image and an OFF-image from tens of minutes (in an in-situ static experimental setup) to less than a millisecond – a reduction by over five orders of magnitude. This mostly overcame the signal variation issues described above and helped us isolate weak signals that were associated with different states of the material/device. Details of the measurement technique are further discussed in the following sections and in the supplementary material.[22] This technique has similarities with prior pump-probe experiments using similar x-ray characterization tools,[23,24] but has the unique provision to perform adaptive switching, discussed in the following section.

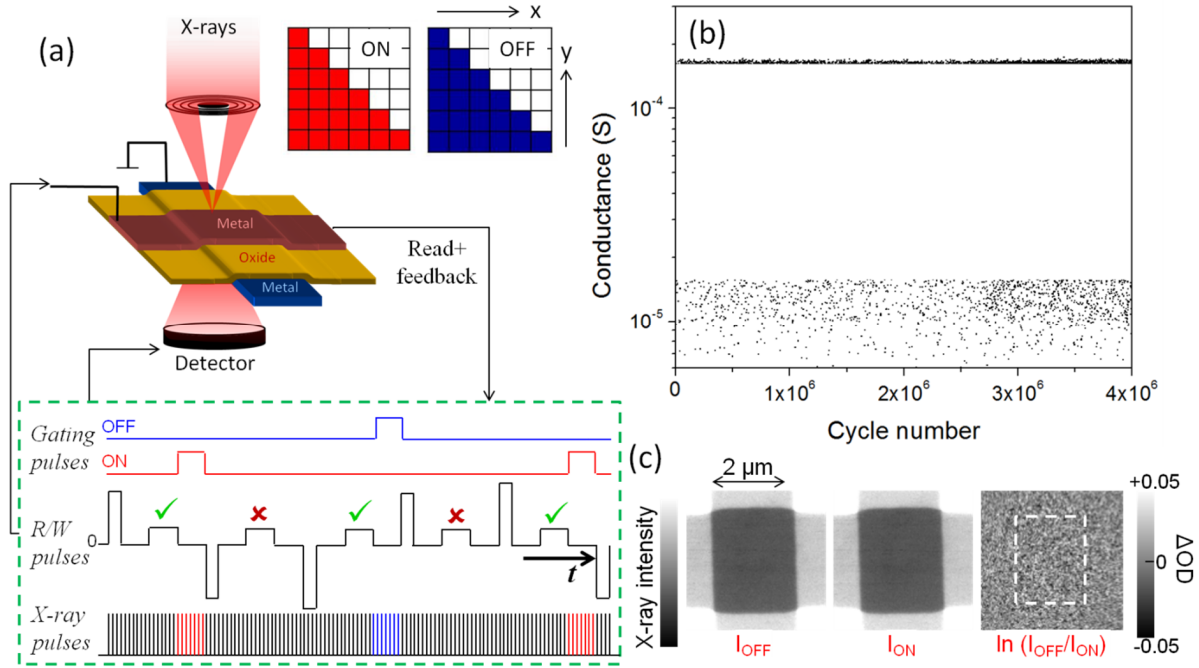

Figure 1: (a) Schematic of the time-multiplexed x-ray absorption spectromicroscopy setup for threshold conductance adaptive switching. The electrical pulses corresponding to read/write (R/W) signal, gating signal are shown along with a representation of the x-ray pulses. Blue and red colors of the x-ray pulses refer to those used to count in different counters gated with color coded gating pulses. Square grids labeled ON and OFF represent parallel acquisition of maps of two different states of the device. (b) Typical threshold conductance (plotted using conductance) switching plot for 4 million successful ON-OFF cycles using adaptive feedback switching as described in the text. Data was sub-sampled for ease of display. (c) Spatial maps over the crosspoint region of the ON and OFF states obtained using the technique summarized in panel (a) at 532.3 eV along with their optical density (OD) difference map. Region within dashed white line is well within the crosspoint region.

## III. THREE MODES OF OPERATION

### A. Bistable Conductance States

Figure 1b is typical data from repeatedly cycling a $TaO_x$ device millions of times between ON-OFF states with a conductance ratio of about 10. In this experiment, the two counters of the gate signal were used to distinguish only the terminal conductance states of the device, following the application of voltage pulses (Figure 1a). By reading the state after every pulse, x-ray mapping of the material in either of the states was performed only when the device had been switched above an ON-conductance threshold or below an OFF-conductance threshold, which was how the ON/OFF ratio was enforced. This allows for an adaptive measurement, gating the x-ray measurement pulses based on the results of the read state. Our measurement approach assumes that repeatedly accessed ON and OFF states have similar material signatures and by summing over many of them the signal accumulates.

In this mode, we used adaptive measurements of memristors (depicted in Figure 1a), where we measure the conductance of the device after every writing pulse to determine if the device switched ('✓') beyond a predefined ON (lower limit) and OFF (upper limit) threshold conductance. In case the device failed ('✘') to reach the threshold conductance, the switching hardware alters the applied voltage pulses (usually increases in amplitude upto <10 V) and attempts repeatedly until the threshold has been reached or crossed. Only after the device is measured to be beyond the ON or OFF threshold is the detector signal gated into either of the two different counters mentioned in the preceding paragraph. Using custom-built hardware we achieved complete software defined control of the input and measurement systems. Thus, this setup could be of generic utility to several different systems with measurable multiple states of operation that can be accessed using synchronous external physical stimuli, e.g. voltage pulses.

Figure 1c displays an ON-state and an OFF-state image obtained using the techniques described above at the energy of the peak of the lowest conduction band (532.3 eV; representative spectra corresponding to the material in these memristors are shown in Figure 2c). Also displayed is the difference in the optical density between the two images. There is no clear localized change visible; although there are some spots that might look different between the images, those are mostly in the same order as the noise. We repeated this exercise of obtaining images with longer dwell times at multiple energies corresponding to key spectral features, especially the $e_g$ and $t_{2g}$ bands (data shown in Figure S2).[22,25,26] We were unable to observe any spatially correlated changes, but in most of the images we saw that the difference between the two states averaged over the entire crosspoint area was consistently different from zero for a given energy. For example, at 532.3 eV (displayed in Figure 1c), the averaged the optical density difference within the crosspoint area was 0.0011±0.0002, and we found this number to be positive irrespective of the spatial region within the crosspoint over which optical density difference was averaged. This was not a constant offset or a similar artifact in our measurement, since the signal level was sufficiently high to enable us to detect changes to the material and such a consistent difference signal was not present outside the crosspoint (discussed further in supplemental material[22]). This indicates that there is a small and uniform (within the spatial resolution) spectral difference between the two conductance states. The signal measured is so low that we were not able to perform spectroscopy in this case (explained in Figure S3).[22] However, based on the spatial-independence observed here, the switching-induced changes of the material may be uniform over the device area or localized channels could be created and destroyed with each ON and OFF pulse in regions distributed across the area.

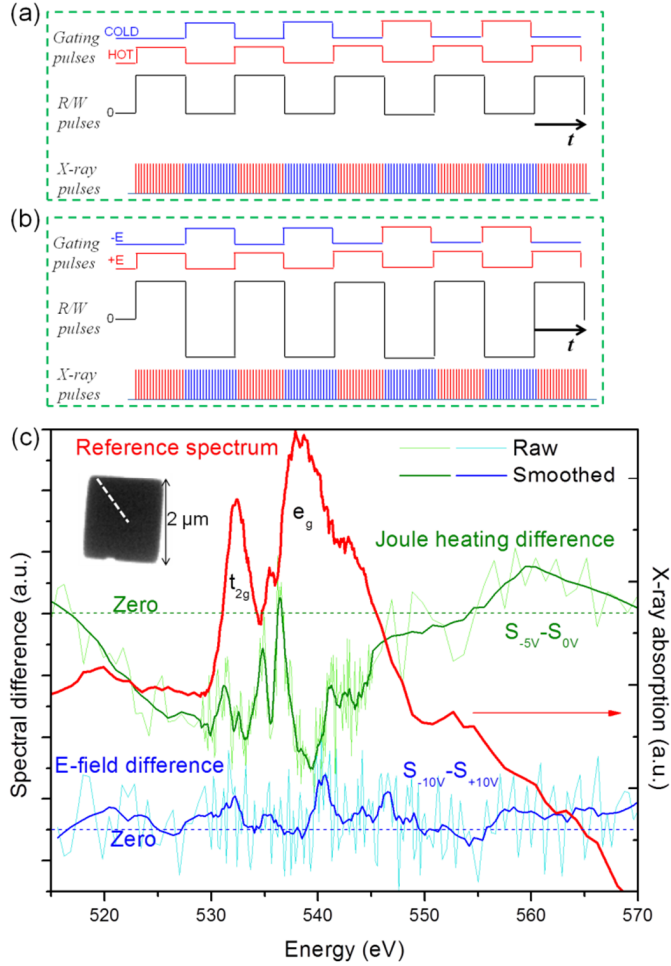

Figure 2: Schematics of (a) Joule heating cycling and (b) Electric field cycling. Blue and red colors of the x-ray pulses refer to those used to count in different counters gated with color coded gating pulses. 'R/W pulses' stands for 'read/write pulses'. (c) A spatially averaged reference spectrum of the material within the crosspoint displayed with the difference between its spectrum in the "hot" state (using -5 V pulses) and that in the "cold" state (using 0 V pulses) and the difference between its spectrum with -10 V across it and that with +10 V across it. The spatial region (along a straight line) over which this data was obtained is shown in the inset, which is an x-ray absorption reference map of the crosspoint. Savitzky-Golay smoothing over 6 data points was used to generate the smoothed spectra. The dashed line at each difference spectra corresponds to its zero (color coded), while they were arbitrarily offset along y-axis for clarity.

**B. Effect of Joule Heating**

We demonstrated the device response to Joule heating by comparing spectromicroscopy of the crosspoint area with and without an applied bias in the high conductance ON state using the time-multiplexed technique described above. The rationale was that high-conductance channels might respond to Joule heating with a spectral signature, while other parts of the crosspoint where relatively little current flowed should have no contrast. We confirmed that these effects were not field driven by comparing similar effects from positive and negative voltage pulses. The input pulses were fixed in amplitude (with equal duration), 5 V (with power levels of ~4 mW) (hot) and 0 V (cold) pulses, and detectors counted only when the pulses were being applied to the device (Figure 2a). Figure 2c shows a reference spectrum of the material along with a difference spectrum between the hot and cold state of the material averaged over a region within the crosspoint (the component hot and cold spectra are displayed in Figure S4).[22] In the difference spectrum, the sharp crests and troughs over the wide background correspond in width and position to the rising and falling edges of the component bands of the spectra (possible origins of the wide background are discussed in the supplemental material[22]). This suggests that the component bands are shifting in energy to produce this form of a difference function, which points to electronic and chemical changes to the "hot" state of the material. In the "hot" state, the lowest conduction band ($t_{2g}$) shifts to lower energies, which may indicate the creation of states in the bandgap and an increase in the conductivity. The $e_g$ band has a significant shift to lower energies, which may result from the Joule heating volume expansion of the material (lengthening of the chemical bonds). As was the case with conductance switching, the spectral signatures were uniformly distributed across the entire crosspoint area (Figure S5).[22] This indicated that channels or regions of highest conductance were spread across the entire crosspoint upon the last ON-switching event, while the conductance state of the material did not change during this data acquisition.

**C. Electric Field Driven Effects**

Since an electric field is hypothesized to cause charged ion and/or vacancy drift that is at least partly responsible for memristive switching in tantalum oxide, a means to visualize the spatial distribution of changes due to an electric field is extremely valuable. To enable this mode, the detector counted only during the application of alternating positive and negative voltage pulses of equal duration (Figure 2b). To increase the signal, we applied fixed external voltage pulses with alternating polarities that were higher (2.5-10 V) than those required to switch the device (1-2 V). Using 10 V pulses with alternating polarities, we studied the spectral changes averaged over a region of the crosspoint (Figure 2c). We were able to see repeatable and distinguishable features in the difference spectra, especially prominent features at about 540 eV and 547 eV and a smaller feature at the lowest conduction band. As was the case with conductance switching and Joule heating, we saw that the spectral signatures were uniformly distributed across the entire crosspoint (supplemental material[22]), thus reinforcing the spatial distribution of changes reported in the previous sections. The signal levels in this case was in the order of 1% in the difference spectrum (Figure S6). Given that the features were barely discernible from the noise, this level of sensitivity for the signal is likely the limit of this technique.

**IV. HIGH VOLTAGE CYCLING**

**A. Observation of Phase Agglomeration**

After the device was cycled over a million times in the Joule-heating and electric-field imaging experiments, the spatial maps of the crosspoint area showed distinct ~100 nm diameter dark and somewhat smaller bright spots (Figure 3a). This was followed by the bright regions appearing in clusters of comparable size as that of the dark regions, while some of the smaller bright regions in the map for ~7x10$^6$ cycles appeared to have agglomerated into larger bright regions after ~6x10$^7$ cycles. We measured the spectra for these regions (Figure 3b) and indeed saw higher absorption in the dark regions compared to the bright regions in the post-edge region, indicating higher elemental concentration of Oxygen in the dark regions. The lowest conduction bands in the bright regions show a downshift by about 0.38 eV relative to those in the dark regions. Also, the lowest conduction band ($t_{2g}$) had a higher density of states relative to the $e_g$ band in the bright regions than in the dark regions. The intermediate intensity, or grey regions, that form most of the background in the images were consistently between the bright and dark in these spectral parameters. Our spectroscopy results indicate that the bright regions are oxygen deficient (~-14%) with a higher conductivity, consistent with an oxygen vacancy agglomeration; and the dark regions are oxygen rich (~+13%) with a lower conductivity.

The spectral feature that appears at 535.4 eV (indicated by a black arrow in Figure 3b) is nearly absent in the bright regions but is very prominent in the dark regions, while less prominent in the grey regions. The disappearance of such a prominent spectral feature indicates dramatic material changes when correlated with the difference in oxygen content. A similar feature at this relative spectral position has been associated[27-32] with a superoxide species, which is an oxygen-oxygen single bond with a net single negative charge that weakly binds to an electropositive species (tantalum) by charge transfer (also argued to be adsorption). This observation provides further insight into the chemical composition of the dark spots in the x-ray images. We correlate these with excess oxygen, as apparent from the references above, and hypothesize the excess oxygen to be interstitial oxygen atoms in tantalum oxide.

**B. Correlation of Device Behavior to Phase Agglomeration**

Since the changes within the crosspoint area x-ray images in response to conventional conductance switching, Joule heating and electric field contrast were all homogeneous, any conductance channels that formed must have been smaller than the spatial resolution and uniformly distributed. When the device was cycled using higher voltages than conventional switching, the power (thus the internal temperature resulting from Joule heating) and the total energy input to the device increased dramatically. Hence, the oxygen vacancies and interstitials in $TaO_x$ can migrate faster and farther than they do during conventional conductance switching, which provides insight into the very long time behavior of a device (a form of accelerated aging of the device).

We were able to image the dark and bright spots for tens of minutes without observing any motion or intereaction of the oxygen rich and deficient agglomerates with each other, despite an expected opposite charge, and thus they have reached a significant level of metastability.[33,34] We were also not able to turn the device OFF (to a low conductance state) after $7 \times 10^6$ high voltage cycles, consistent with the formation of a large irreversible agglomeration of vacancies, i.e. a metallic channel that shorted the device. In contrast, cycling these devices using only conventional conductance switching voltage pulses (1-2 V, shown in Figure S7)[22] allows for a lifetime over $10^8$ cycles without the observation of any such agglomerates. It would be interesting as a future experiment to study the effect of such agglomerates on the surface morphology and their exact chemistry and structure.

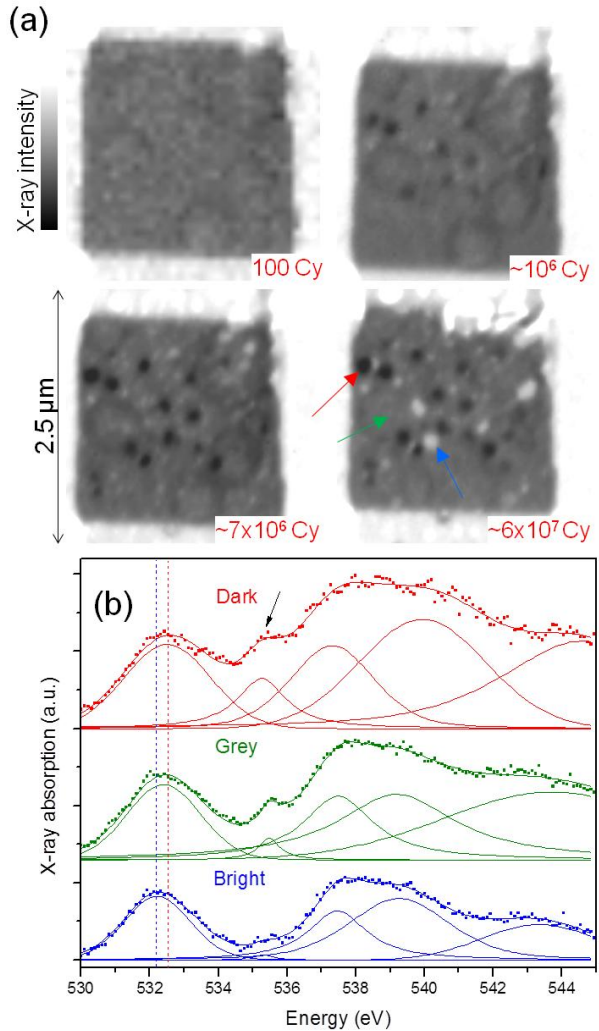

Figure 3: (a) X-ray absorption images of a tantalum oxide memristor at different extents of cycling. All maps were obtained between 530 and 533 eV. (b) O K-edge spectra obtained within the dark, grey and bright regions seen as the device was cycled (panel (a)). Arrows in panel (a) with colors corresponding to that of spectra in panel (b) are used to point out the different regions. The spectra were normalized to a common background, and a linear background in the pre-edge was subtracted from each spectrum. Component bands of each spectrum are shown. The raw spectra are shown in supplemental material (Figure S9).[22]

**V. RELAXATION OBSERVED USING VARYING PULSE WIDTHS**

To better understand the response of the material to electric field, we used an identical configuration as described in the electric field mode, and varied the width of the applied pulses on a fresh device. We hence effectively varied the number of pulses within a pixel of a spatial map without altering the time for which a given voltage (+5 V or -5 V) was applied within the pixel. We observed that as more pulses were packed within a pixel, the field driven effect was stronger irrespective of the region within the crosspoint that was chosen (Figure 4a), further establishing the uniform nature of switching. To discount artifacts, we integrated within different spatial regions and confirmed that no such signal was observed outside of the crosspoint. Given that we apply voltage pulses of hundreds of microseconds, this observation is unlikely to be a transient effect.

A voltage pulse exerts a force on the ions and perturbs the oxide electronic structure, due to both electric field and thermal energy from the Joule heat. This allows the system to overcome energy barriers to find more stable states with time, allowing the ions to relax to a lower energy configuration. Hence, in this case, we believe that the wider pulses allowed for a greater degree of relaxation of the system, closer to its initial spectroscopic properties. Packing higher number of narrower pulses within a time allows the x-rays to average over more changes and lesser relaxation, which is described by Figure 4a. The initial local electronic changes due to the voltage pulse can be seen as abrupt changes to the macroscopic conductance, owing to percolative conduction. But the subsequent relaxation suggested here may not have an impact on the macroscopic conductance unless the extent/nature of relaxation affects the percolation pathways. Irrespective of the conductance, all changes discussed here affect the x-ray spectroscopic signatures since it measures a local property. In other terms, we are isolating the effect of the differential coefficient of the applied energy relative to time. We emphasize that it was the time-multiplexed technique we developed that allowed for this effect to become apparent.

## VI. SPATIAL LOCALIZATION OF SPECTRAL EFFECTS BY ENGINEERING THE FILM

The nature of the oxide film may determine the extent of localization of the conduction. We believe that our grown oxide could have been very uniform in topography, composition and defect distribution, thus giving no favored switching location and allowing for a uniform spread across the entire crosspoint. In one experiment, we used a device with an identical geometry and switching oxide as the one used in this report, except that the lithography on the electrodes allowed for metal to stick up at the edges (by about 100 nm, Figure S8),[22] hence causing topographic distortion. In this case, we performed an experiment identical to the one used in this report to look for Joule heating driven effects and we saw that the effects were localized to the edges of the electrodes and did not show up everywhere inside the crosspoint (Figure 4b) while the spectral signatures and signal levels of the effects remained nearly identical to that seen in this study. In another example, we used a device with similar geometry to the one used here but with a tantalum oxide that was deposited using a co-sputtering technique that was known to be less homogeneous in terms of interface defect states, surface morphology and composition. This is usually intentionally done to create favorable "easy" switching spots. In this device, when we cycled it between ON and OFF conductance states using DC bias, we found a localized region change in oxygen absorption (Figure 4c), although we have yet to eliminate many artifacts before attributing the change purely to the switching event (supplemental material, Figure S1).[22] This suggests that the film can be engineered to localize the switching, but the primary films we used here were homogeneous and displayed a relatively uniform conductance during low-power switching. We emphasize that the localized changes were observed above background noise that was atleast as high as most of the data obtained in this report. This further supports the supposition that the switching could have indeed been uniform in nature and that we did not miss localized non-uniformly spread out filaments.

Interestingly, when the switching was localized, we found that the endurance of the device to repeated cycling decreased especially in devices represented by Figures 4b and 4c (Figure S8e).[22] Here we showed that there are structural changes to the material that are driven by Joule heating, which when localized (concentrated) to a single filamentary region for every cycle of operation, weakens the region and also the films in the stack that interface with it. This leads to early irreversible damage (failure) to the material in that region. This is an important understanding from this work and tells us that making films smoother and devoid of compositional and large topographic inhomogeneities helps in spreading out the switching among many localized filaments and improves the endurance of the devices.

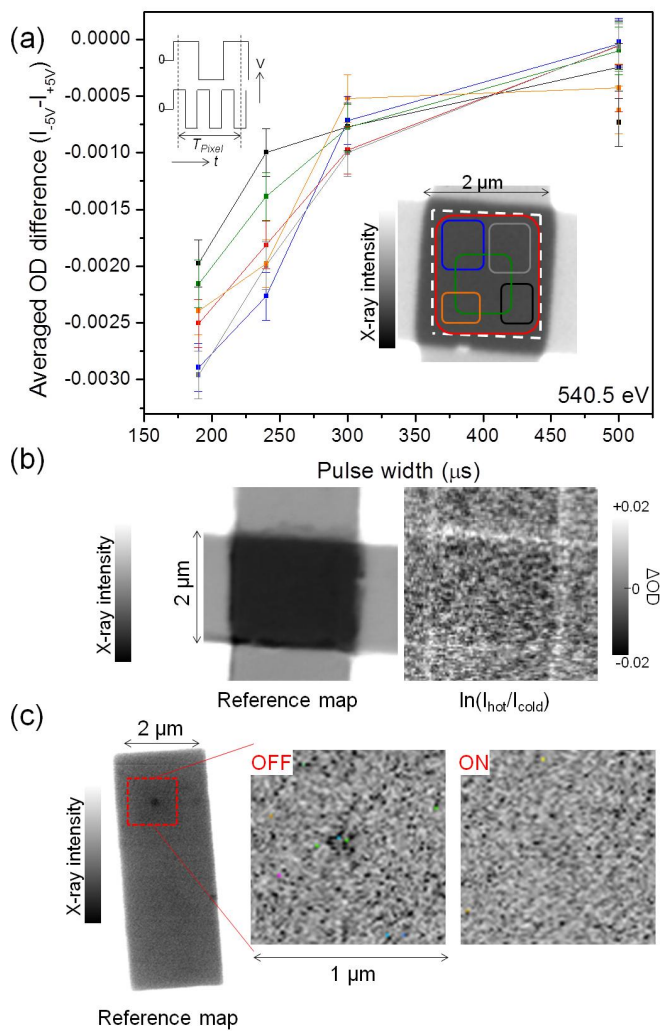

Figure 4: (a) Averaged optical density difference within different regions of a crosspoint (color coded) between the device exposed to -5 V and to +5 V at 540.5 eV. The x-axis is a variation in the pulse width of the applied pulses. All images used to generate this data were averaged for identical time periods at every pixel ($T_{Pixel}$), while the symmetric positive and negative pulses together occupied nearly 100% of the time while their width was varied. A cartoon inset describes this. The experiment was repeated to confirm that the changes with pulse width were not permanent. (b) Joule heating effect seen in a device represented by the reference map obtained using the time-multiplexed technique at 530.75 eV (rising edge of lowest conduction band). 6.5V and 0 V pulses were used for this experiment with devices identical to the ones used throughout this paper except that the edges of the electrodes were allowed to stick up during lithography (SEM images of these devices are shown in Figure S8).[22] The effect can be seen highly pronounced at the edges of the electrode. (c) In-situ static O K-edge mapping of the ON and OFF conductance states of $TaO_x$ device where $TaO_x$ was deposited using a different technique than the one

used throughout this report. The entire crosspoint was mapped in a static ON state and then the device was turned OFF using DC bias after which the crosspoint was mapped in a static OFF state. No dynamic switching or time-multiplexing was used. ON and OFF images were obtained in a smaller spatial region within the crosspoint where we thought there was a change with ON and OFF switching. Details of acquisition are given in the supplemental material.[22]

## VII. CONCLUDING REMARKS

We obtained spatial and spectral x-ray measurements to understand the role of Joule heating and electric fields in inducing electronic and structural changes to tantalum oxide thin films that enable memristance. The observed effects of electrical stimuli were homogenously spread across the entire crosspoint area of our devices; no conducting channels larger than the 30 nm resolution of the images were observed nor were there any other spatial variations in the images across a micrometer scale crosspoint. Upon application of high amplitude voltage pulses (larger than for conventional conductance switching) for millions of cycles, we observed agglomeration of oxygen rich regions followed by that of oxygen deficient regions with continued cycling of the device. This corresponds to the formation of metastable concentrations of oxygen interstitials and conducting channels of oxygen vacancies. The creation of the highly conductive regions corresponded to a very high conductance state, or short, that could not be switched back to a low conductance state, thus highlighting a failure mechanism. We showed that the potential used to drive the device into different states favors relaxation of the material towards its initial state if the energy from the stimuli is injected for longer times. In addition, we showed that the changes to stimuli observed above can be spatially localized by engineering the films' topography and uniformity, which affects their endurance to repeated cycling. This set of observations establishes the capabilities of the time-multiplexed x-ray spectromicroscopy technique introduced here, and also aids in improved understanding of the operation and material evolution in an oxide based memristive device.

## ACKNOWLEDGMENTS

X-ray measurements were performed at the Advanced Light Source at Lawrence Berkeley National Laboratory, CA, USA, at beamlines 5.3.2.2 and 11.0.2. The Advanced Light Source is supported by the Director, Office of Science, Office of Basic Energy Sciences, of the U.S. Department of Energy under Contract No. DE-AC02-05CH11231.

## REFERENCES

[1]L. Chua, Circuit Theory, IEEE Transactions on **18**, 507 (1971).
[2]T. Prodromakis, C. Toumazou, and L. Chua, Nature materials **11**, 478 (2012).
[3]J. Borghetti, G. S. Snider, P. J. Kuekes, J. J. Yang, D. R. Stewart, and R. S. Williams, Nature **464**, 873 (2010).
[4]J. J. Yang, D. B. Strukov, and D. R. Stewart, Nature nanotechnology **8**, 13 (2013).
[5]M. D. Pickett, G. Medeiros-Ribeiro, and R. S. Williams, Nature materials **12**, 114 (2012).
[6]A. Rua, R. Cabrera, H. Coy, E. Merced, N. Sepulveda, and F. E. Fernandez, Journal of Applied Physics **111**, 10 (2012).
[7]M. A. Kats, D. Sharma, J. Lin, P. Genevet, R. Blanchard, Z. Yang, M. M. Qazilbash, D. N. Basov, S. Ramanathan, and F. Capasso, Applied Physics Letters **101** (2012).
[8]B. Hu, Y. Ding, W. Chen, D. Kulkarni, Y. Shen, V. V. Tsukruk, and Z. L. Wang, Advanced Materials **22**, 5134 (2010).

[9] F. Miao, J. P. Strachan, J. J. Yang, M. X. Zhang, I. Goldfarb, A. C. Torrezan, P. Eschbach, R. D. Kelley, G. Medeiros-Ribeiro, and R. S. Williams, Advanced Materials **23**, 5633 (2011).

[10] J. J. Yang, M. X. Zhang, M. D. Pickett, F. Miao, J. P. Strachan, W.-D. Li, W. Yi, D. A. A. Ohlberg, B. J. Choi, and W. Wu, Applied Physics Letters **100**, 113501 (2012).

[11] J. J. Yang, M. X. Zhang, J. P. Strachan, F. Miao, M. D. Pickett, R. D. Kelley, G. Medeiros-Ribeiro, and R. S. Williams, Applied Physics Letters **97**, 232102 (2010).

[12] A. C. Torrezan, J. P. Strachan, G. Medeiros-Ribeiro, and R. S. Williams, Nanotechnology **22**, 485203 (2011).

[13] H. S. Wong, H.-Y. Lee, S. Yu, Y.-S. Chen, Y. Wu, P.-S. Chen, B. Lee, F. T. Chen, and M.-J. Tsai, Proceedings of the IEEE **100**, 1951 (2012).

[14] I. Goldfarb, F. Miao, J. J. Yang, W. Yi, J. P. Strachan, M. X. Zhang, M. D. Pickett, G. Medeiros-Ribeiro, and R. S. Williams, Applied Physics a-Materials Science & Processing **107**, 1 (2012).

[15] J. P. Strachan, A. C. Torrezan, G. Medeiros-Ribeiro, and R. S. Williams, Nanotechnology **22**, 505402 (2011).

[16] J. P. Strachan, G. Medeiros-Ribeiro, J. J. Yang, M. X. Zhang, F. Miao, I. Goldfarb, M. Holt, V. Rose, and R. S. Williams, Applied Physics Letters **98**, 242114 (2011).

[17] K. Kamiya, M. Y. Yang, T. Nagata, S.-G. Park, B. Magyari-Köpe, T. Chikyow, K. Yamada, M. Niwa, Y. Nishi, and K. Shiraishi, Physical Review B **87**, 155201 (2013).

[18] A. L. D. Kilcoyne, T. Tyliszczak, W. F. Steele, S. Fakra, P. Hitchcock, K. Franck, E. Anderson, B. Harteneck, E. G. Rightor, G. E. Mitchell, A. P. Hitchcock, L. Yang, T. Warwick, and H. Ade, Journal of Synchrotron Radiation **10**, 125 (2003).

[19] W. Chao, P. Fischer, T. Tyliszczak, S. Rekawa, E. Anderson, and P. Naulleau, Optics Express **20**, 9777 (2012).

[20] L. Dongyi, Z. Yadong, A. Tran Xuan, Y. Yu Hong, H. Daming, L. Yinyin, D. Shi-Jin, W. Peng-Fei, and L. Ming-Fu, Electron Devices, IEEE Transactions on **61**, 2294 (2014).

[21] J. P. Strachan, M. D. Pickett, J. J. Yang, S. Aloni, A. L. David Kilcoyne, G. Medeiros-Ribeiro, and R. Stanley Williams, Advanced Materials **22**, 3573 (2010).

[22] See supplemental material for additional data and explanations that support this manuscript.

[23] D. P. Bernstein, B. Bräuer, R. Kukreja, J. Stöhr, T. Hauet, J. Cucchiara, S. Mangin, J. A. Katine, T. Tyliszczak, K. W. Chou, and Y. Acremann, Physical Review B **83**, 180410 (2011).

[24] Y. Acremann, J. P. Strachan, V. Chembrolu, S. D. Andrews, T. Tyliszczak, J. A. Katine, M. J. Carey, B. M. Clemens, H. C. Siegmann, and J. Stöhr, Physical Review Letters **96**, 217202 (2006).

[25] L. Soriano, M. Abbate, J. Vogel, J. C. Fuggle, A. Fernández, A. R. González-Elipe, M. Sacchi, and J. M. Sanz, Surface Science **290**, 427 (1993).

[26] L. Soriano, M. Abbate, J. C. Fuggle, M. A. Jimenez, J. M. Sanz, C. Mythen, and H. A. Padmore, Solid state communications **87**, 699 (1993).

[27] M. W. Ruckman, J. Chen, S. L. Qiu, P. Kuiper, M. Strongin, and B. I. Dunlap, Physical review letters **67**, 2533 (1991).

[28] S. L. Qiu, C. L. Lin, J. Chen, and M. Strongin, Journal of Vacuum Science & Technology A **8**, 2595 (1990).

[29] S. L. Qiu, C. L. Lin, J. Chen, and M. Strongin, Physical Review B **41**, 7467 (1990).

[30] A. Knop-Gericke, M. Hävecker, T. Schedel-Niedrig, and R. Schlögl, Topics in Catalysis **10**, 187 (2000).

[31] A. Dilks, Journal of Polymer Science: Polymer Chemistry Edition **19**, 1319 (1981).

[32] J. G. Chen, B. Frühberger, and M. L. Colaianni, Journal of Vacuum Science & Technology A **14**, 1668 (1996).

[33] B. Magyari-Köpe, S. G. Park, H.-D. Lee, and Y. Nishi, Journal of Materials Science **47**, 7498 (2012).

[34] G. Grimvall, B. Magyari-Köpe, V. Ozoliņš, and K. A. Persson, Reviews of Modern Physics **84**, 945 (2012).

# SUPPLEMENTARY MATERIAL

**Processing details**

The silicon nitride membranes on which the devices were built were made by a wet-etching process. A 200 nm film of silicon nitride was grown on both sides of a double-side polished 250 μm thick silicon wafer using low pressure chemical vapor deposition to give a silicon nitride film with <250 MPa tensile stress. Silicon on one side of the wafer was exposed by removing the nitride layer with photolithography + reactive ion etching, which was followed by etching in ~25% KOH for 6-8 hours to etch the window through the silicon wafer. This was then cleaned using the RCA process followed by a vacuum bake to remove all ionic impurities, metallic particles, oxide impurities and moisture, before fabricating the device over the window. The device stack was 3 nm Ti (bottom electrode) + 20 nm Pt (bottom electrode) + 6 nm $TaO_x$ (blanket) + 7 nm Ta (top electrode) + 20 nm Pt (top electrode). The electrodes were defined using photolithography and the electrode metals were all deposited by e-beam induced evaporation.

The $TaO_x$ was deposited using ion beam induced sputtered deposition (using an Oxford Instruments Ionfab 300Plus machine): the target was $Ta_2O_5$, base pressure 2.3e-7 torr, processing gas Ar = 4 sccm, $O_2$ = 1 sccm (through ion gun), pressure 2.5 to 2.6e-4 torr, and the deposition time was 72 seconds for 6 nm at room temperature.

Oxide B (or $TaO_x$ B), referenced in Figures S1, S8, was grown using another sputtering tool (using a CHA Industries Solution sputtering machine) using co-sputtering of two targets (Ta and $Ta_2O_5$) Ta 0.4% Ta2O5 125W (RF) Ar = 93sccm. This film was purposely less uniform to encourage the formation of favorable switching regions.

**The problem with in-situ static measurements**

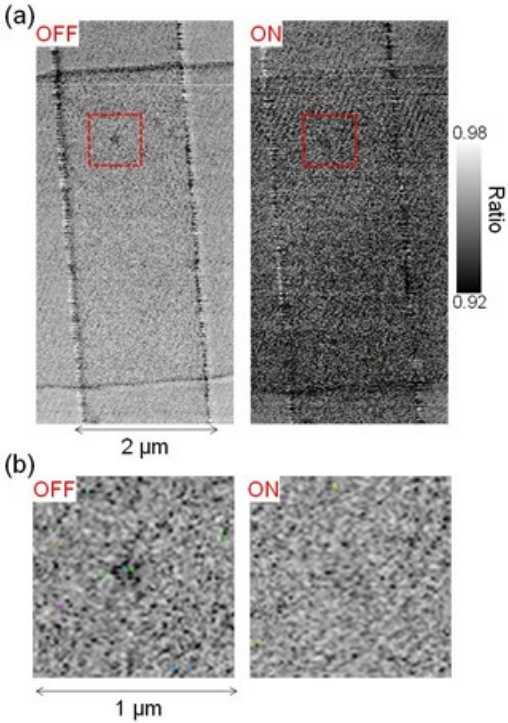

Figure S1: (a) OFF-state and ON-state static images from intensity maps obtained at 531 eV (the rising edge of the lowest conduction band corresponding to the tantalum oxide, or Oxide B, used here) and normalized to a pre-edge map at 529 eV (after correcting for spatial drift) to correct for variation in absorption between acquisitions of the maps. Both images are plotted on the same intensity scale. (b) Cropped and magnified parts of the two images shown in panel (a) within the dashed red squares. These maps are plotted on slightly different intensity scales to allow the differences to become more apparent – especially a dark spot in the OFF-state image which is less pronounced in the ON-state image.

From Figure S1a, we can clearly see that despite correcting for spatial drift using multiple algorithms, we still do not have good registration between the two maps (map at 531 eV and that at 529 eV). This is obvious because we see the edges of the crosspoint in the ratio maps. The bad registration is due to the minutes to tens of minutes of time it takes to obtain a single map, during which there is intra-map drift that is difficult to correct. In addition, despite normalizing to the pre-edge map in each case, we still see continuous variation in intensity in the ratio maps from top to bottom of each map, which is the change in background absorption that happens within a single map. These two artifacts are nearly impossible to eliminate if we hope to get good signals out of small effects. Because of these artifacts, there is a hanging question on whether the changes that we observe in the localized spot shown in Figure S1b can be attributed to the ON-OFF switching. This is the reason we implement and advocate for the time-multiplexed technique used throughout the report.

**Further description of the lock-in x-ray method and modes of operation**

As mentioned in the text, the time-multiplexed x-ray spectromicroscopy method was used in three modes: threshold resistance switching, joule heating cycling, and electric field cycling. For threshold resistance switching, the voltage pulse width was 2 µs, with a 10 µs delay between the write and read voltages, while the read voltage was 0.2 V. The write voltages were usually between 1 V and 2.5 V for an ON-OFF conductance ratio of 10 (see Figure S7 for an example x-ray image with these conditions). There were approximately 1-2 successful write events out of about 4 attempts within a millisecond. This gives us a write-pulse-coverage of only 4 μs within a millisecond. In the threshold resistance switching mode, as shown in Figure 1a, the x-ray pulses are used to read the state of the device only after the read cycle confirms that the device is in the required state and the voltage across the device is zero so this mode looks purely at the differences between the resistance states.

In the joule heating mode (Figure 2a), gating is done depending on whether or not a pulse is being applied to the device. Equal durations of a positive or negative voltage pulse or zero voltage are applied to the device to make it alternately "hot" or "cold". In this mode, we did not continuously read the resistance state of the device (which is an overhead on time) or use feedback. We preferred to use negative voltages on the top electrode.

The electric field cycling (Figure 2b) was identical to the joule heating driven mode, except that instead of a finite potential and zero potential repeating in time, we used finite potentials alternating in polarity. We chose to use fields higher than that required to switch the device between threshold resistance states, i.e. to switch the device with an ON-OFF conductance ratio of 10, we usually required 1-2.5 V, whereas in the electric field cycling mode, we used voltage amplitudes of 2.5-10 V, mostly to enhance the signal. Due to the high fields involved, conductance switching was at times not observable; most of the devices usually measured a low resistance state despite multiple high voltage pulses of alternating polarity. This was probably due to several strong conductance channels formed by the high amplitude voltage pulses, which effectively shorted the devices and prevented OFF-switching. Despite this, we expected to see some field-dependent changes in the material.

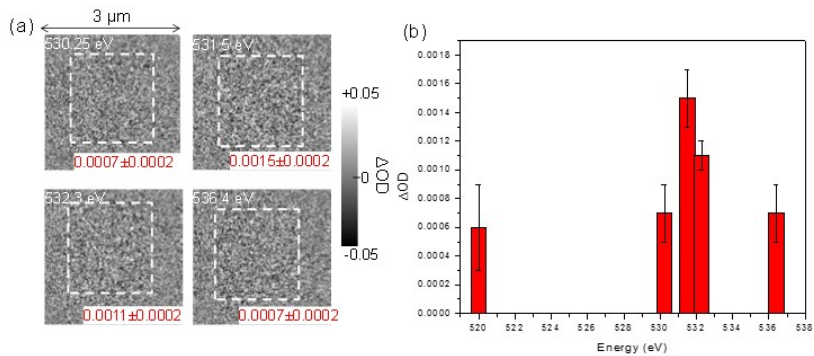

Figure S2: (a) Spatial x-ray absorption maps showing the natural logarithm of the ratio of the OFF and ON maps at different x-ray energies (in the threshold conductance switching mode) and hence representing the change in optical density (OD) between the two conductance states. A dashed square is included to indicate the approximate position of the crosspoint device. All maps were obtained in the same spatial region covering the entire crosspoint area. In each of the maps, the average intensity of all pixels within the crosspoint was a finite number other than zero, indicated in each map, along with the uncertainty, the root mean square of the noise envelope divided by the square root of the total number of pixels used to obtain the reported signal. (b) The change in OD within a crosspoint at different energies represented by a vertical bar plot, with the uncertainties or errors included.

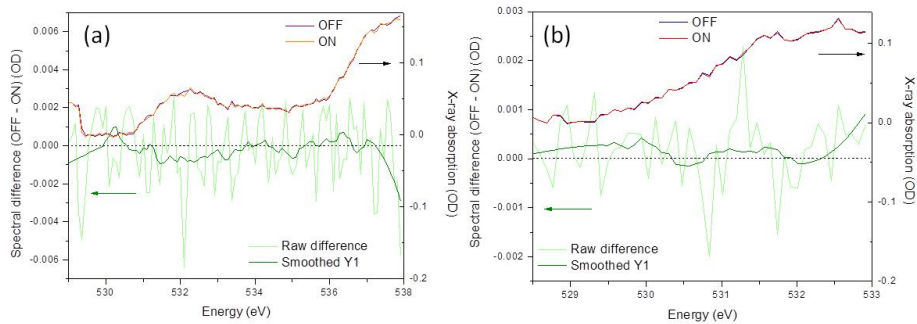

Figure S3: Spectra corresponding to the ON and OFF conductance states of the device. (b) was obtained in a part of the spectral region of (a). The panels show that there is not enough signal to isolate clear features in the difference spectra. This fact is clearer when the signal levels here are compared to the signal level observed in the maps shown in Figure S2.

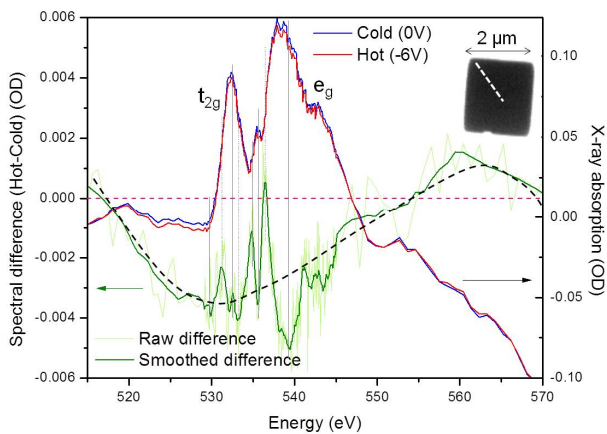

Figure S4: The component x-ray absorption spectra corresponding to the joule heating difference spectrum shown in Figure 2c. The difference spectrum is displayed again to enable comparison. Inset shows the region in which a line-scan was performed (also shown in the inset of Figure 2c). The green dashed line following the difference spectrum is a free hand sketch to indicate the broad background. This helps isolate the features corresponding to the absorption peaks. We did not display the component hot and cold spectra in the main text because, to the naked eye, there is no appreciable difference between the two, unless one is subtracted from the other to highlight the differences.

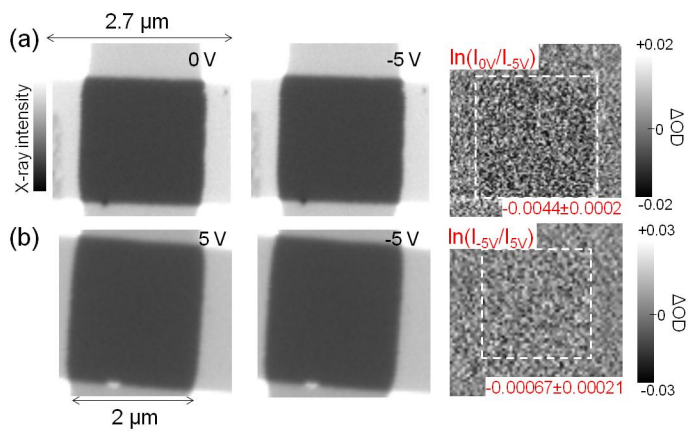

Figure S5: (a) Joule heating effect on a crosspoint device studied by mapping the rising edge of the lowest conduction band (531.5 eV) along with the change in optical density map (ln(hot/cold)). The averaged signal inside the crosspoint is -0.0044±0.0002, which corresponds well with the signal level in the difference spectrum at 531.5 eV (roughly -0.004) in Figure S4, which was obtained using similar electrical stimuli. (b) Electric field effect studied on the same device as panel (a) at 540.5 eV along with the change in optical density map (ln(positive filed/negative field)). The averaged signal within the crosspoint was -0.00067±0.00021. This corresponds to about half the signal level (roughly 0.0015) in the difference spectrum in Figure S6, which was obtained using twice the field amplitude used in this measurement.

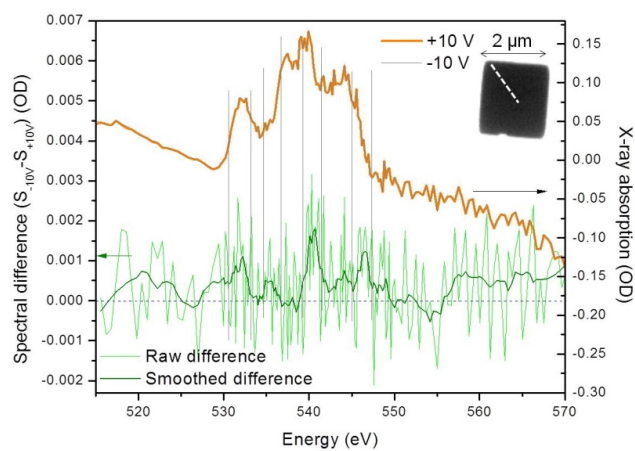

Figure S6: The component spectra corresponding to the electric field difference spectrum shown in Figure 2c. The difference spectrum is displayed again to enable comparison. Inset shows the region in which a line-scan was performed (also shown in the inset of Figure 2c). The spectra in this plot have more noise, mainly due to the noise in the background spectrum that was used to normalize the absolute absorption.

**Possible artifacts in Joule heating effects**

The broad background seen in the Joule heating difference spectrum in Figure 2c could be pulse-induced spatial drift caused by thermal flexing of the membrane holding the device (Figure S12). It is also possible that the broad background was due to hot electrons during the "hot" cycle in the Pt N-edge signal which is expected to appear in a similar spectral position.

We considered possible reasons why a nanoscale filament was not spectrally discernible in the Joule heating mode. First, it is possible that the current density through the whole device (including the filament but also surrounding areas) was so high such that spectral changes were occurring a multiple spatial points in the sample. But the effect seemed to scale with applied current or voltage and persisted even when the applied voltages were less than that required to switch the resistance states of the device. We also considered the possibility that the signal was not high enough and there was some strange offset between the counters of the two states. But as mentioned in the preceding section, such an effect was absent outside the crosspoint area, and can therefore be excluded. Also, we have observed similar spectral changes with similar signal levels in localized regions in devices that were slightly different from the ones used here (Figure 4b).

There is an artifact that we have not been able to discount yet. It was argued previously in the main text that the spatially uniform signals during threshold resistance switching indicated two possibilities: that the switching could indeed be uniform or spread out in nature; or every on-switching event produces a new filament in a new spatial location and completely annihilates or renders it inactive upon off-switching, such that a map averaged over many switching events appears to show spatially uniform or smeared out changes. Looking at the joule heating signals in a low resistance device, without altering its resistance state, was intended as a trick used to locate the conductive channels that were created during the previous ON-switching event.

Now that we see uniform changes with joule heating, we might conclude that the ON-switching event indeed induced a uniform change. But the potential used to perform the joule heating experiment was 5 V, while that used for threshold resistance switching was 1-2 V. This is a considerably larger power. As we have seen in the other parts of the report, a higher power or energy could cause dramatic changes to the behavior of the material. Without elaborating on possible alternate outcomes, ideally, we should have used 1-2 V to unambiguously confirm the effect, but we did not see any contrast at all with 1-2 V.

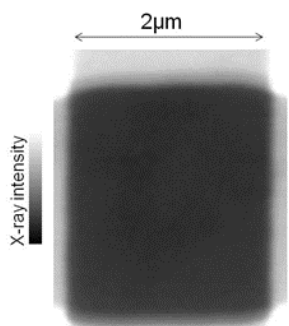

Figure S7: X-ray absorption (at 532.3 eV) map of a device that had undergone ~$10^8$ threshold conductance switching cycles. Map was obtained after the cycling was done. Unlike the case of the device in Figure 3, there are no clear agglomerates of oxygen rich or oxygen deficient regions. This is linked to the fact that there is far less energy pumped into the device in the threshold resistance switching mode compared to the joule heating or electric field cycling modes. This is because of both pulse amplitude and the percentage of time that is covered by the pulses, apparent from Figures 1a, 2.

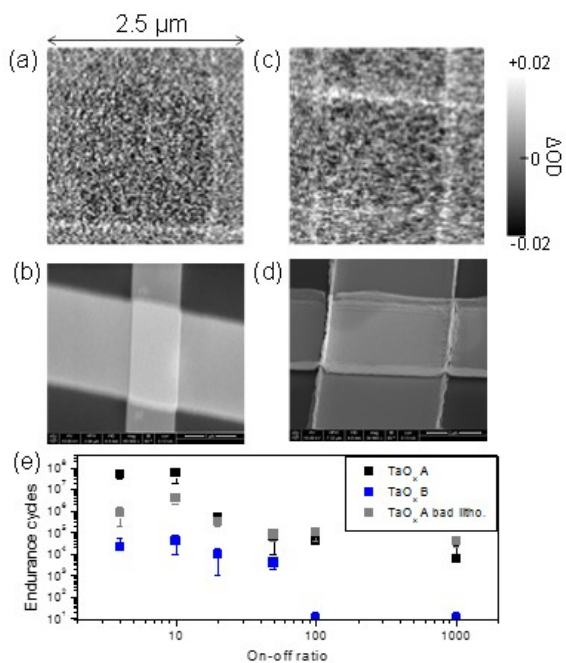

Figure S8: (a) is a reproduction of the optical density difference map in Figure S5a. (b) is an SEM image of a device made in the same batch on the same chip as that shown in (a). (c) is a reproduction of the difference map shown in Figure 4b. (d) is an SEM image of a similar device made in the same batch on the same chip as that shown in (b). It is fairly clear that having metal edges on the crosspoint sticking up helps in localizing the effects of joule heating, probably due to field enhancement and lower resistance offered for current flow in these regions. This is to aid in interpreting Figures 4b-4c; changes to the material can be localized in the film's topography or the composition can be modified. We confirmed that the changes in both (a) and (c) show nearly identical spectral behavior, which indicates that this is purely a physical localization of the same effect. (e) Endurance of devices made of $TaO_x$ (the oxide used throughout this study), $TaO_x$ with bad lithography correspond to device represented by (d). $TaO_x$ B corresponds to the device represented by Figure 4c. Y-axis represents endurance of devices to repeated threshold resistance switching, while the x-axis is the on-off ratio that is enforced onto the devices through the adaptive characterization mode shown in Figure 1a.

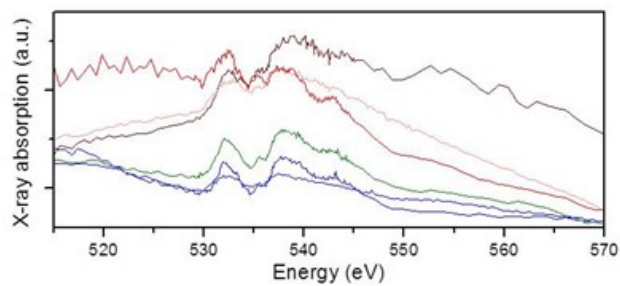

Figure S9: Raw spectra obtained from the dark (red), grey (green) and bright (blue) regions in Figure 3a. The spectra were normalized to a common background absorption. Spectra in Figure 3b consist of three of the spectra shown here, from the three different regions, and they underwent a pre-edge slope correction and were pinned to zero at the pre-edge.

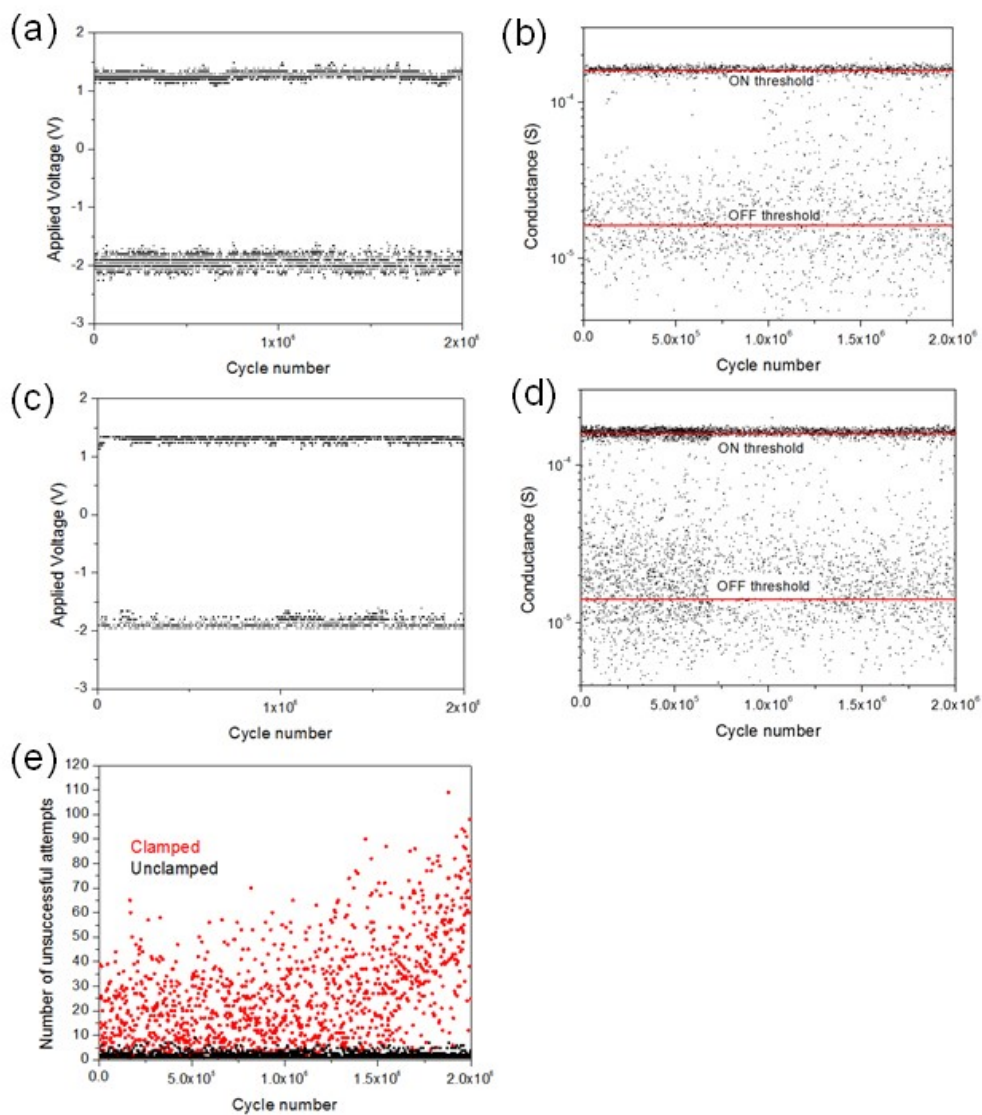

Figure S10: (a) is input voltage pulse amplitudes for threshold resistance switching that corresponds to a conductance distribution with successful switching cycles as shown in panel (b). In another threshold resistance switching experiment, the maximum allowed voltage pulse amplitude was clamped at half the voltage distribution in both the positive and negative voltages, for the same device, resulting in a voltage pulse amplitude distribution shown in panel (c) corresponding to a conductance distribution shown in panel (d). The solid red lines are the ON-state and OFF-state thresholds used in the adaptive resistance switching mode. Panel (e) is a plot of the number of unsuccessful attempts for every successful switching cycle for both the clamped case and unclamped case.

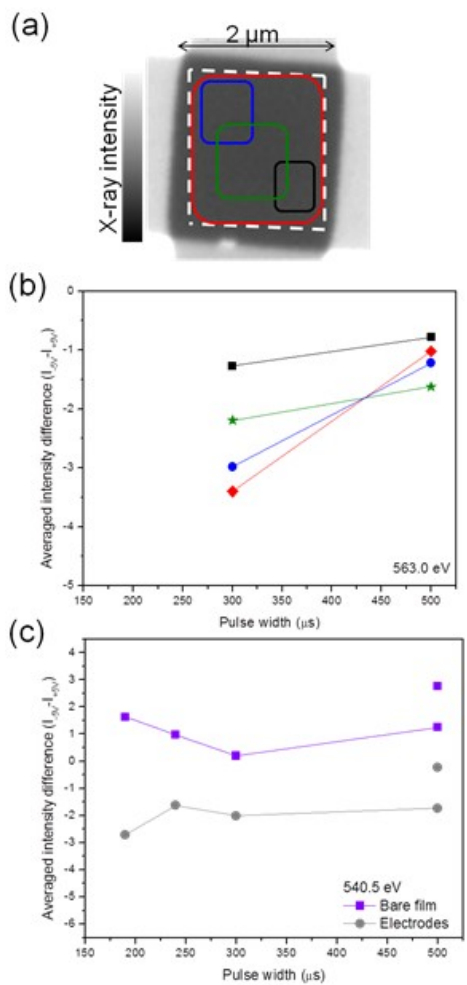

Figure S11: Electric field dependence of the optical density difference with varying pulse width. Panel (a) is a part of the inset in Figure 4a. (b) Identical plot as that of Figure 4a, except for different x-ray energies used, as noted in the plot. (c) Identical plot as that of Figure 4a, except that the averaged spatial region is outside the crosspoint, as noted in the legend. This plot was made to show that there is no pulse width dependence unlike in Figure 4a or Figure S11b outside the crosspoint. This pulse width experiment was done on a fresh device and we used lower voltages than before (5 V) so that we do not "damage" the device like the one seen in Figure 3a before the end of measurement.

**Calculation of the number of O atoms from x-ray data**

We estimate that the sensitivity of our detection scheme was a change of less than 4000 oxygen atoms per scan spot (about 30 nm x 30 nm, or < 1 $nm^{-3}$) averaged over ~200 switching cycles, with dwell time per pixel of 200 ms. All calculations on elemental oxygen concentrations were done by assuming that the post-edge optical densities represented the elemental concentration through Beer-Lambert's law. The spectra considered for this calculation were all normalized to zero in the pre-edge to account for any non-oxygen changes.

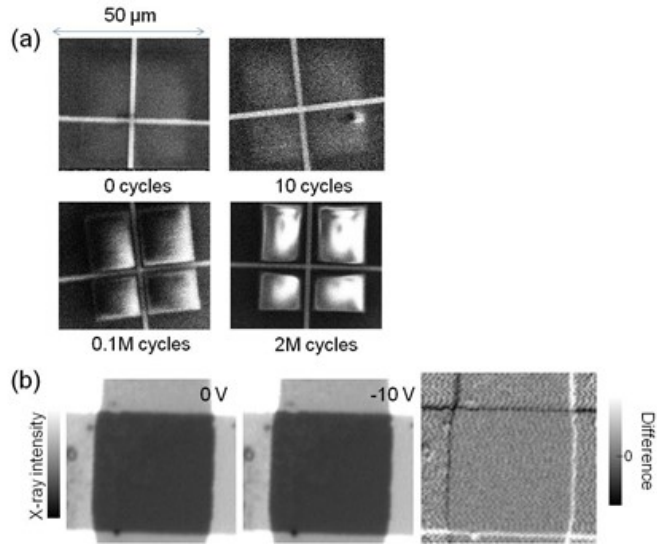

Figure S12: (a) SEM images of different devices, with each device having been cycled (in threshold resistance switching mode) to different extents as noted below each image. The devices are built on top of 200 nm thick silicon nitride membranes to allow for x-ray transmission. The membrane (about 40 μm x 40 μm) is seen in these images, and it appears to have deformed under stress in each of the 4 quadrants, divided by the electrodes. The electrodes are the only good heat sinks in proximity to the devices, so most of the heat that is dissipated into the membrane ends up stressing it, causing irreversible deformation. This is evidence for the membrane flexing with joule heating. (b) X-ray absorption maps at 536.1 eV to investigate joule heating cycling. The difference map (hot – cold) is also shown, while the component maps are marked for the voltage pulse amplitude applied. The dark or bright outlines on the electrodes occur due to spatial drift in a dominant direction. In this case, since it is occurring repeatedly, we know that it has to be because of joule heating. We believe that joule heating could be flexing the membrane hence causing a drift and potential defocusing of the hot image.